\documentclass[aps,prl,footinbib,twocolumn,floatfix,superscriptaddress]{revtex4}
\usepackage{amsfonts,amssymb,epsfig,verbatim,mathbbol,amstext,amsmath,psfrag}

\def\P{\mathcal{P}}
\def\tr{\mbox{tr}}

\def\Nc{N}
\def\Ns{{N_s}}

\def\eq{Eqn.}
\def\eqs{Eqns.}

\newcommand{\mo}[1]{(#1)_{\!\!\!\!\!\mod 1}}

\begin{document}

\title{A random matrix-like model for the Polyakov loop and center symmetry}

\author{Falk Bruckmann}

\affiliation{Institut f\"ur Theoretische Physik, Universit\"at Regensburg, D-93040 Regensburg, Germany.}

\begin{abstract}
We formulate a random matrix-like model for the Polyakov loop in $SU(\Nc)$ Yang-Mills theories. It describes a simplified dynamics in terms of eigenvalue differences. The deconfinement phase transition encoded in center symmetry breaking is reproduced including its nature being first order for $SU(3)$ and second order for $SU(2)$. Analytical arguments about the phases are presented and a comparison to other approaches is made.
\end{abstract}


\maketitle

\emph{Introduction} ---
Evidence for the quark-gluon plasma state from heavy ion experiments has triggered plenty of theoretical and numerical work on the phase diagram of Quantum Chromodynamics (QCD). Deconfinement and restoration of chiral symmetry are the principal phenomena to be described at the phase transition. As intrinsically nonperturbative effects they remain difficult to derive from first principles. In perturbation theory, for instance, quarks are not confined.

Random matrix theory -- in contrast to ab initio-methods like e.g.\ lattice gauge theory -- governs universal spectral properties. In QCD this usually concerns the Dirac operator. Through the density of Dirac eigenvalues at zero eigenvalue, random matrix theories contain a chiral condensate quite naturally. Furthermore, temperature can be built into the formalism via the Mat\-su\-ba\-ra frequencies  (see \cite{Verbaarschot:2000} and references therein). Although such random matrix models are schematic, they correctly incorporate the chiral restoration at high temperatures. In this way random matrix models help to understand one of the principal phenomena of QCD, in a very simple system with almost all QCD dynamics removed.

In this Letter we ask the question whether in the same spirit one can investigate the finite temperature deconfinement transition, too. The Polyakov loop is related to the free energy of (infinitely heavy) quarks and hence the deconfinement order parameter at finite temperature. It can be studied in the quenched approximation aka Yang-Mills theories.
In the latter case, the Polyakov loop is a strict order parameter as it respects/breaks center symmetry at low/high temperatures, respectively (see below).

We shall introduce a random matrix-inspired  model for the Polyakov loop eigenvalues and their effective potential depending on temperature. As argued this is achieved already in pure Yang-Mills theories, with general gauge group $SU(\Nc)$. Our model reveals center phase transitions with the correct order for $SU(2)$ and $SU(3)$. Thus also center symmetry breaking and hence deconfinement can be understood by virtue of a simplified model based on random matrices.\\

\emph{The Polyakov loop and center symmetry}  --- 
The Polyakov loop as an element of the gauge group $SU(\Nc)$ can be diagonalized to
\begin{equation}
 \P=\exp\big(2\pi i \, \mbox{diag}(\mu_1,\mu_2,\ldots,\mu_\Nc)\big)
\end{equation}
where the parameters $\mu_a$ are dimensionless numbers and sum to an integer. Of course, the $\mu$'s are defined only up to permutations and adding integers, which will be discussed in more detail below.

The average over the trace 
\begin{equation}
L\equiv \frac{1}{\Nc}\,\tr\,\P=\frac{1}{\Nc}\sum_{a=1}^{\Nc} \exp(2\pi i \mu_a)
\label{eqn L def}
\end{equation}
is an order parameter for (de)confinement via center symmetry \cite{Polyakov:1978_Susskind:1979} (see \cite{Pisarski:2009} for a recent review):
The Polyakov loop is traceless, $\langle L\rangle=0$, below the critical temperature,
which is achieved by \emph{equidistant eigenvalues}
\begin{eqnarray}
 \mu_1=\Delta\mu,\:\mu_2=\Delta\mu+\frac{1}{\Nc},\ldots,
\mu_\Nc=\Delta\mu+\frac{\Nc-1}{\Nc}\,,
\end{eqnarray}
where an additional shift $\Delta\mu=1/2\Nc$ is needed for even $\Nc$ to ensure $\det\P=1$. 

At high temperatures the absolute value of the Polyakov becomes maximal, $\langle|L|\rangle\to 1$, which amounts to \emph{degenerate eigenvalues}
\begin{equation}
 \mu_1,\:\mu_2,\:\mu_3,\ldots,\mu_\Nc \to \mu\,.
\end{equation}
Here $\mu$ is 0 for real Polyakov loop $L\to 1$, in the quenched case at high temperature $\mu$ also approaches (with same probability) multiples of $1/\Nc$ with $P$ being a nontrivial center element of $SU(\Nc)$ and $L$ being a nontrivial $\Nc$th root of unity.

This reflects center symmetry, which is the invariance under transformations
\begin{equation}
L\to\exp(2\pi i k/\Nc)L\,,\qquad \mu_a\to\mu_a+\frac{k}{\Nc}
\end{equation}
with integer $k$.\\ 

\emph{``Deriving'' the model} ---
As is natural for random matrix models, we will consider constant fields. The temporal gauge field $A_0$ needs to reproduce the Polyakov loop,
\begin{equation}
A_0=\frac{1}{i\beta}\log \P= 2\pi T\,\mbox{diag}(\mu_1,\mu_2,\ldots,\mu_\Nc)\,.
\label{eqn A0 start}
\end{equation}
This choice amounts to the Polyakov gauge whose Faddeev-Popov determinant (Jacobian) is known to be the reduced Haar measure \cite{Weiss:1981_Gross:1981_Weiss:1982}
\begin{equation}
 \prod_{a > b}\sin^2\big(\pi(\mu_a-\mu_b)\big)\,.
\end{equation}
This kinetic term gives sets of eigenvalues $\mu_a$ different weights according to the associated volume of gauge orbits.

In our model the spatial gauge fields $A_i$ will take the role of fluctuations and hence are replaced by hermitian matrices $X_i$. 

The action proportional to $F_{\mu\nu}^2=\sum_{\mu,\nu}\tr(i[A_\mu,A_\nu])^2$ is treated as follows:
The magnetic part $-\sum_{i,j}\tr[A_i,A_j]^2$ is replaced by the quadratic $\sum_i \tr\, X_i^2$, Gaussian weights for the random matrices $X_i$. Assuming universality, the precise form of the weight for $X_i$ should not influence the results much. 

The electric part $-\sum_i\tr[A_0,A_i]^2$ is quadratic in both $A_i$ and $A_0$. Upon replacing the gauge fields by the matrices $X_i$ and the rhs. of \eq~(\ref{eqn A0 start}), respectively, this part consists of $2\times 2$-blocks
\begin{equation}
 -\tr\Big[\left(\begin{array}{cc}\mu_a&\\&\mu_b\end{array}\right)\!,\!
\left(\begin{array}{cc}X_i^{aa}&X_i^{ab}\\X_i^{ab\,*}&X_i^{bb}\end{array}\right)\Big]^2=
2\,(\mu_a-\mu_b)^2 |X_i^{ab}|^2
\label{eqn block}
\end{equation} 
to be multiplied by $(2\pi T)^2$. 

Here the ambiguity in the $\mu$'s leads to a subtlety. Consider momentarily ordered $\mu$'s according to 
(e.g.~\cite{Kraan:1998a_Diakonov:2007})
$\mu_1\leq \mu_2 \leq\ldots\leq\mu_\Nc\leq \mu_1+1\cong\mu_1$
where the equivalence $\cong$ stands for the same eigenvalue $\exp(2\pi i \mu_a)$ in the Polyakov loop.
Then for two indices $a>b$:
\begin{equation}
\mu_b\leq \mu_a \leq\mu_b+1\cong \mu_b\,.
\end{equation} 

The block considered so far, \eq~(\ref{eqn block}), takes into account the commutator of the spatial gauge field with $A_0\sim\mbox{diag}(\mu_a,\mu_b)$ resulting in a term with the difference $\mu_a-\mu_b$. However, $A_0\sim\mbox{diag}(\mu_a,\mu_b+1)$ is equally valid and so is the difference $\mu_b+1-\mu_a=1-(\mu_a-\mu_b)$.
The above inequalities limit these differences to the interval $[0,1]$ and it depends on the configuration which one is shorter (yielding a smaller action).

For symmetry reasons we therefore include also the other difference \footnote{One could also include higher differences such as $\mu_a+2-\mu_b$, but we want to keep the model as simple as possible.}, 
through a term
\begin{equation}
\big(1-(\mu_a-\mu_b)\big)^2 |X_i^{\prime\,ab}|^2
\end{equation} 
with new fluctuation matrices $X'_i$ subject to the same quadratic action $\sum_i\tr\, X_i'^2$. 

Both $X$ and $X'$ are $\Nc\!\times\!\Nc$-matrices for gauge group $SU(\Nc)$ (unlike in conventional random matrix models, where the limit of infinite size is performed).

As our model depends only on the differences of $\mu$'s center symmetry is manifest. We will abbreviate the difference of two $\mu$'s by $\nu$. The ordering of the eigenvalues can be taken into account by the definition
\begin{equation}
\nu_{ab}\equiv\mo{\mu_a-\mu_b}\in[0,1)
\end{equation} 
in terms of which we can finally write down our model:
\begin{eqnarray}
&&Z(\{\mu\})\equiv\prod_{a > b}\sin^2\big(\pi\nu_{ab}\big)\label{eqn the model}\\
&&\times \prod_{i}\int\! dX_i dX_i' 
\exp\!\Big(\!\!-(2\pi C)^2 \tr (X_i^2 +X_i'^2)\Big)\nonumber
\\
&&\times\exp\!\Big(\!\!-\sum_{a > b}2(2\pi T)^2\big[\nu_{ab}^2|X_i^{ab}|^2+
(1-\nu_{ab})^2|X_i^{\prime\,ab}|^2\big]\Big)\nonumber
\end{eqnarray}
thereby introducing the parameter $C$, a `coupling constant' of the model.

The effective potential for the Polyakov loop eigenvalues is related to $Z$ by $Z=\exp(-V/T)$, hence up to $\nu$-independent constants
\begin{eqnarray}
 \frac{V(\{\mu\})}{T}&=&\sum_{a > b}\Big\{-\log\sin^2\big(\pi\nu_{ab}\big)\label{eqn V}\\ 
&&+\Ns \log\Big(\frac{C^2}{T^2}+\nu_{ab}^2\Big)
\Big(\frac{C^2}{T^2}+(1-\nu_{ab})^2\Big)\Big\}\,,\nonumber
\end{eqnarray}
with $\Ns$ the number of spatial dimensions ($i=1,\ldots,\Ns$ in (\ref{eqn the model})), in due course put to 3 unless specified otherwise.\\

\begin{figure}[b]
\includegraphics[width=0.9\linewidth]{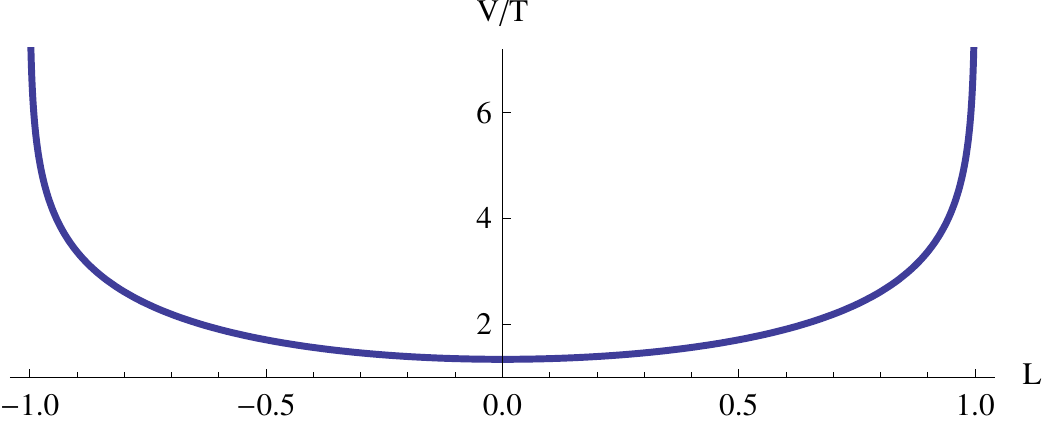}\\
\vspace{0.3cm}
\includegraphics[width=0.9\linewidth]{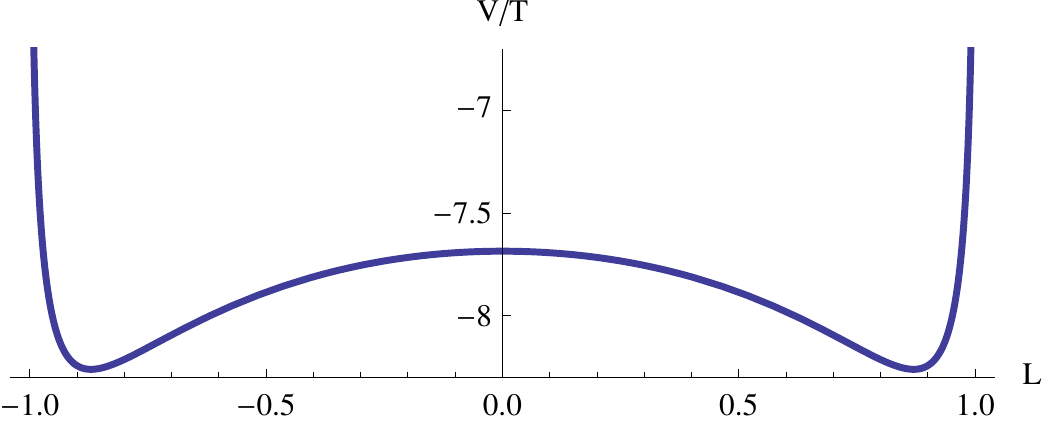}
\caption{Effective potential $V/T$ over the traced Polyakov loop $L$ for gauge group $SU(2)$,
\eq~(\protect\ref{eqn V L}) (obtained from \eqs~(\protect\ref{eqn V}) and (\protect\ref{eqn L def})), at low temperature $T/C=1.0$ (upper panel) and high temperature $T/C=6.0$ (lower panel), respectively.}
\label{fig su2 overview}
\end{figure}

\emph{Results and interpretation} ---
In the dynamics of Polyakov loop eigenvalues the Haar measure term suppresses degenerate eigenvalues, for which $\nu_{ab}=0$, and hence tends to confine, as is known from strong coupling expansion on the lattice, see \cite{Fukushima:2004} and references therein.

The $\nu_{ab}^2$-term in the second line of \eq~(\ref{eqn V}) prefers nearby eigenvalues, which however renders the $(1-\nu_{ab})^2$-term large (for $\mu_a=\mu_b+\epsilon$, the terms act vice versa for $\mu_a=\mu_b-\epsilon$). Which $\mu$ configurations eventually yield the minimal potential $V$ then depends on the accompanying factors and hence on temperature measured in units of~$C$.

In Fig.~\ref{fig su2 overview} we show for two temperatures the effective potential in $SU(2)$ over the traced Polyakov loop $L$ as given by
\begin{eqnarray}
&&\frac{V\{\mu\}}{2T}=-\log(1-L^2)\label{eqn V L}\\ 
&&+3 \log\Big(\frac{C^2}{T^2}+\big(\frac{\arccos L}{\pi}\big)^2\Big)\Big(\frac{C^2}{T^2}+\big(1-\frac{\arccos L}{\pi}\big)^2\Big)\,.\nonumber 
\end{eqnarray} 
The minimum of the effective potential moves from $L=0$ at low temperatures to $|L|\to 1$ at higher ones as it should. The invariance under $L \to -L$ reflects the manifest center symmetry of the effective potential.

\begin{figure}[!t]
\includegraphics[width=0.9\linewidth]{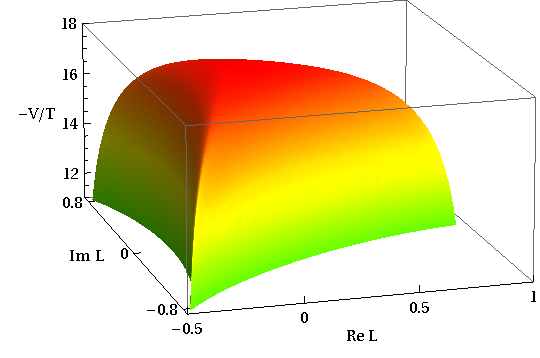}\\
\includegraphics[width=0.9\linewidth]{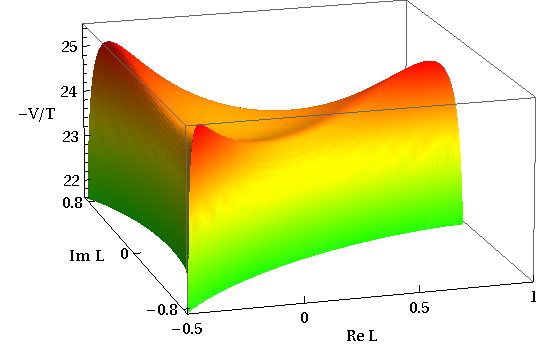}\\
\caption{The \emph{inverse} effective potential $-V/T$ over the traced Polyakov loop $L$ for gauge group $SU(3)$, \eqs~(\protect\ref{eqn V}) and (\protect\ref{eqn L def}), at low temperature $T/C=3.0$ (upper panel)
and high temperature $T/C=6.5$ (lower panel), respectively.}
\label{fig su3 overview}
\end{figure}

The corresponding plots for gauge group $SU(3)$ are shown in Fig.~\ref{fig su3 overview}. Here $V$ is plotted over the complex $L$, whose values are bounded by an approximate triangle. Again, $L=0$ is preferred at low temperatures, whereas at high temperatures the minima are at $L\to \{1,\,\pm \exp(2\pi i/3)\}$ as expected.

We note that for $SU(2)$ and $SU(3)$ there is a one-to-one mapping of Polyakov loop traces $L$ (real resp.\ complex) to configurations of eigenvalues $\mu_a$ (one resp.\ two independent ones). A fixed $L$ in $SU(\Nc)$ with $\Nc\!\geq\! 4$ is generated by different $\mu_a$'s and hence has different potentials $V$ (the minimal one should be taken).\\  

Our model correctly predicts the confining, center-respecting Polyakov loop $L=0$ at low temperatures and a breaking of center symmetry by the `ground state' with $|L|\to 1$ in the high temperature deconfining phase (with $L=0$ turned into a local maximum). 

\begin{figure}[t]
\includegraphics[width=0.89\linewidth]{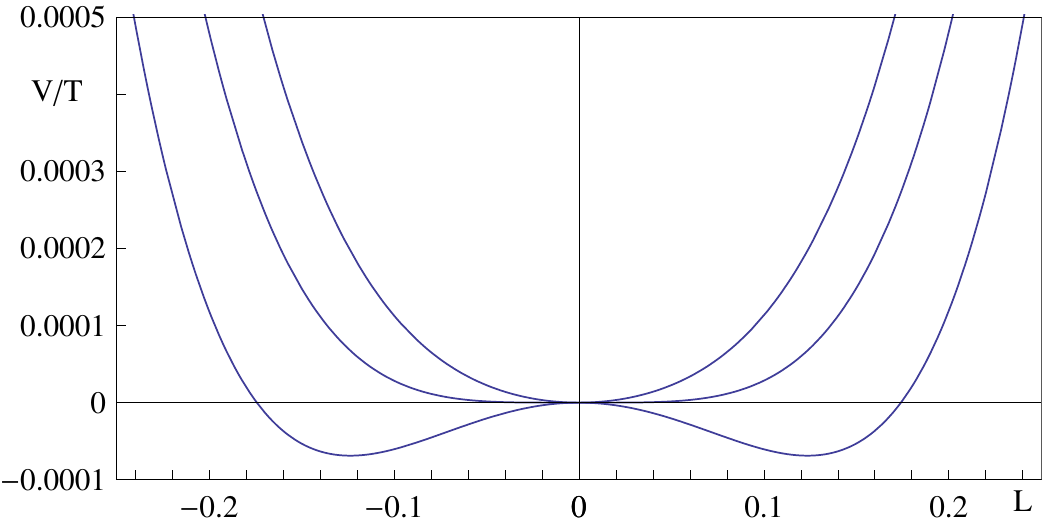}\\
\vspace{0.1cm}
\includegraphics[width=0.89\linewidth]{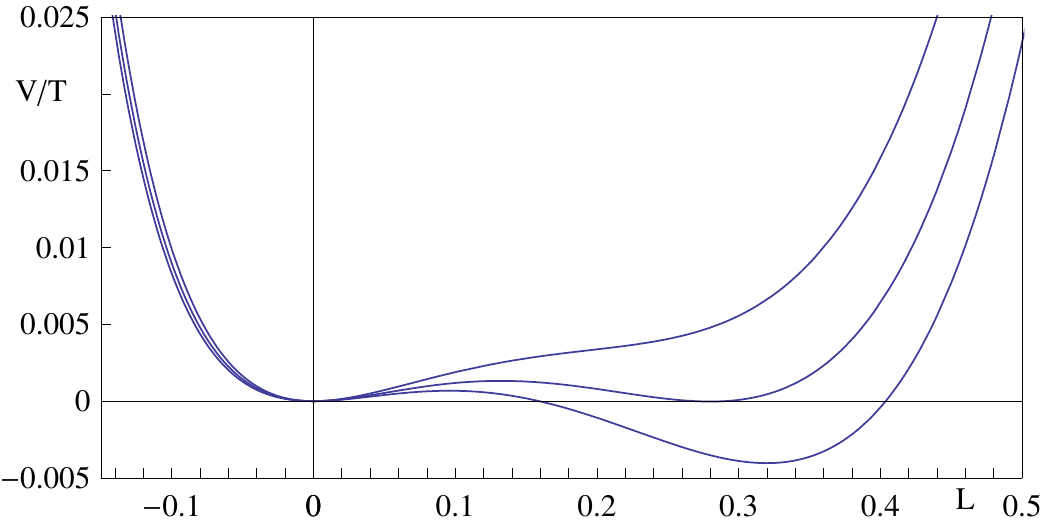}
\caption{Effective potential $V/T$ (with $V/T|_{L=0}$ subtracted) as a function of $L$ zoomed in for three temperatures near the phase transition. Upper panel: $SU(2)$ with $T/C=3.62,\, 3.637,\, 3.655$, a second order transition. Lower panel: real Polyakov loops in $SU(3)$ (parametrizing $\mu_1=-\mu_3\in[0,1/2],\:\mu_2=0,\: L=[1+2\cos(2\pi \mu_1)]/3$) with $T/C=4.53,\, 4.5635,\,4.588$, a first order transition.}
\label{fig order both}
\end{figure}

What is more, also \emph{the order of the transition} in our model is the well-known one for Yang-Mills theories. As can be read off from Fig.~\ref{fig order both}, the phase transition is second order for $SU(2)$ since the finite $L$-minimum develops in a continuous manner out of the $L=0$-minimum. In contrast, the transition is  first order for $SU(3)$, since there is a phase of coexisting local minima, above which the finite $L$-minimum becomes the global one.\\

The low and high temperature behavior of the effective potential can be understood through approximations of the second line of \eq~(\ref{eqn V}). At low $T$ the argument of the log is expanded around the corresponding power of $C^2/T^2$, the first $\nu$-dependent term is proportional to the sum $\sum_{a > b}$ over
\begin{equation}
3\,\frac{T^2}{C^2}\,\big(\nu_{ab}^2+(1-\nu_{ab})^2\big)\,.
\label{eqn low}
\end{equation}
This is maximal at $\nu_{ab}=0$ and thus suppresses degenerate eigenvalues like the Haar measure term. Because of the $T^2$-prefactor in (\ref{eqn low})
the Haar measure is actually the leading term at low temperatures resulting in confining Polyakov loops.

At high $T$ the leading log-term is the sum $\sum_{a > b}$ over
\begin{equation}
3\log\big((\nu_{ab})^2(1-\nu_{ab})^2)\,.
\label{eqn product}
\end{equation}
As the argument is positive it is minimized by vanishing $\nu_{ab}$, hence this term prefers deconfining degenerate eigenvalues. Whether a deconfining Polyakov loop is taken on eventually depends on the dominance of this contribution over the Haar measure $-\log\sin^2(\pi\nu_{ab})$. As is clear from the figures \ref{fig su2 overview} and \ref{fig su3 overview} this is indeed the case for the physical situation of three spatial dimensions. 

For arbitrary number $\Ns$ of spatial dimensions the second $\nu_{ab}$-derivative at intermediate $\nu_{ab}=1/2$ of the term (\ref{eqn product}) and the Haar measure is $-16\Ns$ vs.\ $2\pi^2$, respectively. This suggests a transition to deconfinement at high temperatures for $\Ns\geq 2$. This argument is exact for $SU(2)$ -- where there is just one independent eigenvalue difference $\nu_{21}$ -- but is supposed to hold also for higher gauge groups.

Interestingly, the effective potential at high temperatures, \eq~(\ref{eqn product}), has as argument the one-loop result \cite{Weiss:1981_Gross:1981_Weiss:1982}. It follows that both are minimized by the same $\mu$'s. However, our $V$ is the log of the one-loop expression. There seems to be no reason why the random matrix model built on constant gauge fields could exactly reproduce perturbation theory
or effective Polyakov loop models \cite{Engelhardt:1998_Dumitru:2004_Diakonov:2004b_Wipf:2007}
which are based on varying gauge fields (and typically invoke character expansions).


Phenomenological approaches like Ginzburg-Landau for second order phase transitions or \cite{Straley:1973} for first order ones use low powers of the order parameter near the critical temperature. We make the common ansatz
\begin{eqnarray}
\frac{V_{\rm phen}(L)}{T}=-\frac{b_2}{2}|L|^2-\frac{b_3}{6}(L^3+\bar{L}^3)+\frac{b_4}{4}|L|^4
\label{eqn V phen}
\end{eqnarray}
where the cubic term is utilized  only for $SU(3)$. 

The coefficients $b_{2,3,4}$ can be obtained in $SU(2)$ by Taylor expanding \eq~(\ref{eqn V L}) around $L=0$ to fourth order. In $SU(3)$ we compare the $(\mu_1,\mu_2)$-Taylor series of $V(\mu_1,\mu_2,-\mu_1-\mu_2)$ with $V_{\rm phen}(L(\mu_1,\mu_2,-\mu_1-\mu_2))$ near the confining value $(\mu_1,\mu_2)=(1/3,0)$. 

In our model, $b_{2,3,4}$ come out as rational functions of $T/C$ [not displayed]. In $SU(2)$ the quadratic coefficient $b_2$ turns positive at a critical $T/C=3.637$ (the solution of a certain quartic equation) signalling a second order phase transition, cf.\ Fig.~\ref{fig order both} upper panel. In $SU(3)$ this coefficient turns negative at $T/C=4.718$. The finite $L$-minimum develops at a slightly lower critical temperature $T/C=4.56$, cf.\ Fig.~\ref{fig order both} lower panel.

Around these temperatures the higher coefficients $b_{3,4}$ are 
positive which is expected and e.g.\ agrees qualitatively with the fits given in \cite{Ratti:2006}. While the polynomial approximation to $V$  describes the transition region fairly well, it breaks down at the latest at $T/C\geq 4.88$ resp. $T/C\geq6.51$ where the quartic coefficients for $SU(2)$ resp. $SU(3)$ turn negative and higher powers of $L$ become important.

\emph{Summary and outlook} ---
We have formulated a random-matrix like model for the Polyakov loop eigenvalues, \eq~(\ref{eqn V}) (using continuum language). It contains a deconfinement transition of correct order for $SU(2)$ and $SU(3)$. Through the (symmetrized) interaction with spatial gauge fields the Polyakov loop eigenvalues change from equidistant to nearly degenerate thereby breaking center symmetry. Hence this simple model captures the relevant finite temperature mechanism with only one parameter $C$ setting the temperature scale. 

The underlying space only enters through its dimensionality (the number of spatial gauge fields) hence the model neglects e.g. Polyakov loop clusters and topological objects that might drive the transition.

In the quenched case the model could be used to investigate the phase transition in the large $N$ limit (for the Wilson loop eigenvalues universal properties have been analyzed in this limit \cite{Gross:1980_Durhuus:1981_Brzoska:2005_Narayanan:2007}) and in exotic gauge groups. In $SU(3)$ there are also lattice data for the Polyakov loop in higher representations \cite{Gupta:2008} to compare to.

Furthermore, it should be straightforward to include quarks into the model, along the lines of the Dirac operator in chiral random matrix theory. We conjecture a weakening of the phase transition. In turn, our approach makes it possible to incorporate the Polyakov loop and hence center symmetry in random matrix models. Obviously an extension of the model to nonzero chemical potential would be interesting, too.\\

The author likes to thank Jacques Bloch, Kenji Fukushima, Christof Gattringer, Tilo Wettig, Andreas Wipf and in particular Jac Verbaarschot for many helpful discussions. This work has been supported by DFG (Contract No. BR 2872/4-2).


\end{document}